\documentclass[useAMS,usenatbib]{mn2e}
\usepackage{graphicx, color, setspace, ulem}
\title[Density structure and universalities of DM halos]{The connection between the cusp--to--core transformation and observational universalities of DM halos}
\author[Ogiya, Mori, Ishiyama and Burkert]{Go Ogiya$^{1}$\thanks{E-mail:ogiya@ccs.tsukuba.ac.jp}, Masao Mori$^{1,2}$, Tomoaki Ishiyama$^{2}$ and Andreas Burkert$^{3,4}$\\
$^{1}$Graduate School of Pure and Applied Science, University of Tsukuba, 1-1-1, Tennodai, Tsukuba, Ibaraki, 305-8577, Japan\\
$^{2}$Center for Computational Sciences, University of Tsukuba, 1-1-1, Tennodai, Tsukuba, Ibaraki, 305-8577, Japan \\
$^{3}$Universit\"ats-Sternwarte M\"unchen, Scheinerstra\ss e 1, D-81679 M\"unchen, Germany\\
$^{4}$Max-Planck-Fellow, Max-Planck-Institut f\"ur extraterrestrische Physik, Postfach 1312, Giessenbachstra\ss e, D-85741 Garching, Germany
}
\begin{document}

\date{Accepted 2014 February 03. Received 2014 February 03; in original form 2013 September 06}

\pagerange{\pageref{firstpage}--\pageref{lastpage}} \pubyear{2014}

\maketitle

\label{firstpage}

\vspace{-5mm}
\begin{abstract}
Observations have revealed interesting universal properties of dark matter (DM) halos especially around low-mass galaxies.
Strigari et al. (2008) showed that DM halos have common enclosed masses within 300pc (Strigari relation). 
Kormendy \& Freeman (2004) reported DM halos having almost identical central surface densities (the $\mu_{\rm 0D}$ relation). 
In addition, there exists a core--cusp problem, a discrepancy of the central density distribution between simulated 
halos and observations. 
We investigate whether a scenario where cuspy halos transform into cores by some dynamical processes can also explain their universal structural properties. 
It is shown that a cusp--to--core transformation model naturally reproduces the $\mu_{\rm 0D}$ relation and that Strigari relation follows from the $\mu_{\rm 0D}$ relation for dwarf galaxies. 
We also show that the central densities of cored dark halos provide valuable information about their
formation redshifts.
\end{abstract}


\vspace{-5mm}
\section{Introduction}
The predictions of the present standard paradigm for structure formation in the universe, 
the cold dark matter (CDM) cosmology, excellently match observations on large scales ($\ga {\rm 1Mpc}$). 
However, some serious discrepancies between CDM predictions and observations are being discussed 
on smaller scales ($\la {\rm 1Mpc}$).  One of them is the core--cusp problem, that is the mismatch of 
the observationally inferred central density structures of DM halos when compared with theoretical predictions.
Cosmological CDM $N$--body simulations claim divergent density distributions, called cusps, in the
centers of DM halos (e.g., Navarro et al. 1997, 2010; Fukushige \& Makino 1997; Moore et al. 1999a; Ishiyama et al. 2013). 
The Navarro--Frenk--White (NFW) profile (Navarro et al. 1997) well fits the density structures of simulated CDM halos, 
\begin{equation}
\rho(r) = \rho_{\rm s} r_{\rm s}^3/[r(r+r_{\rm s})^2], \label{nfw}
\end{equation}
where $r$ is the distance from the center and $\rho_{\rm s}$ and $r_{\rm s}$ are the scale density and scale length of the halo, respectively. 
On the other hand, DM halos of observed galaxies, especially for low-mass galaxies, indicate 
constant density profiles, called cores or at least shallower cusps 
(e.g., Moore 1994; de Blok et al. 2001; Swaters et al. 2003; Spekkens et al. 2005; Oh et al. 2011). 
The Burkert profile (Burkert 1995) reproduces well the density structures of DM halos obtained from observation data (see also Salucci \& Burkert 2000), 
\begin{equation}
\rho(r) = \rho_{\rm 0} r_{\rm 0}^3/[(r+r_{\rm 0})(r^2+r_{\rm 0}^2)], \label{burkert}
\end{equation}
where $\rho_{\rm 0}$ and $r_{\rm 0}$ are the central density and the core radius of the halo, respectively.
If DM particles are only interacting gravitationally, flattening out a central DM cusp
would require 
changes in the gravitational potential.
Several mechanisms have been proposed and indeed succeed to reduce the central density.
One of them are changes in the gravitational potential caused by stellar feedback redistributing gas or
generating galactic winds (e.g., Navarro et al. 1996, Ogiya \& Mori 2011, 2012; Pontzen \& Governato 2012: Teyssier et al. 2013). 
Another example is the heating by dynamical friction of massive clumps (e.g., El--Zant et al. 2001; Goerdt et al. 2010; Inoue \& Saitoh 2011). 

In connection with this problem, some studies reported observational universalities of DM halos. 
Strigari et al. (2008) found there exists a common mass scale among DM halos of nearby dwarf galaxies. 
According to their analysis 
based on Jeans equation for spherical systems, the enclosed DM mass within 300pc from the halo centers, $M(<300{\rm pc})$, is $\sim 10^7 M_{\rm \sun}$ (hereafter, Strigari relation). 
Motivated by these observations, Walker et al. (2009) argued that all dwarfs are embedded in the same universal DM halo.
Hayashi \& Chiba (2012) evaluated $M(<300{\rm pc})$ using non--spherical Jeans analysis. 
Independent of Strigari relation, Kormendy \& Freeman (2004) revealed the quantity, $\mu_{\rm 0D} \equiv \rho_{\rm 0} r_{\rm 0}$, 
which corresponds to the central surface density of DM halos for cored profiles, does not depend on blue magnitude of galaxies (hereafter, the $\mu_{\rm 0D}$ relation). 
Donato et al. (2009) confirmed this result for a larger sample. 
The finding of Kormendy \& Freeman (2004) is also consistent with the observed correlations of core parameters
$\rho_{\rm 0} \propto r_{\rm 0}^{-1}$ when fitting the observations by Burkert halos as confirmed by Salucci et al. (2012). 
They also obtained $\mu_{\rm 0D} \sim 140 M_{\rm \sun} {\rm pc^{-2}}$. 
Using the recent observational data, Cardone \& Tortora (2010), Boyarsky et al. (2010), and Cardone \& Del Popolo (2012) argued for a weak dependence of a halo mass on the $\mu_{\rm 0D}$ relation. More recently, Del Popolo et al. (2013) proposed the theoretical interpretation.

The motivation of this work is to investigate the connection between processes transforming cusps to cores and the universalities of dark halo cores
. The outline of this $letter$ is as follows. In order to understand the meaning of the universalities in the cosmological context,
we start in Section 2 with an investigation of the dependence of the dark halo concentration paremeter 
on halo mass and redshift. 
In Section 3, we argue the $\mu_{\rm 0D}$ relation and cusp--to--core transformation process are connected. The results derived in Section 3
provide information about halo properties, formation redshift and virial mass. 
Section 4 demonstrates Strigari and the $\mu_{\rm 0D}$ relation for nearby dwarf galaxies are 
linked that also shows both relations are inconsistent for larger and smaller galaxies. 
We summarize the results in Section 5. 

\vspace{-5mm}
\section{Concentration parameter}
The concentration parameter of an NFW halo, $c$, is defined by $c \equiv r_{\rm 200}/r_{\rm s}$, 
where $r_{\rm 200}$ is the radius inside of which the mean density of the DM halo is 200 times 
of the critical density of the universe and which is called the virial radius. 
This is related to the virial mass, $M_{\rm 200}$, by 
\begin{equation}
M_{\rm 200} \equiv (4 \pi / 3) 200 \rho_{\rm crit} (1+z)^3 r_{\rm 200}^3,
\label{virial_mass}
\end{equation}
where $\rho_{\rm crit}$ and $z$ are the critical density of the universe and redshift, respectively.
If we know the dependence of $c$ on $M_{\rm 200}$ and $z$, we can determine the parameters of the NFW halo, 
$\rho_{\rm s}$ and $r_{\rm s}$, for given $M_{\rm 200}$ and $z$. 
Such dependence, $c(M_{\rm 200},z)$, has been investigated by using cosmological 
$N$--body studies (e.g., Bullock et al. 2001; Macci{\`o} et al. 2008), most of which
concluded that $c$ decreases monotonically with increasing $M_{\rm 200}$ and decreasing $z$. This is
a result of the fact that $c$ reflects the cosmic density at the formation (collapse) epoch of DM halos. 
In the standard CDM cosmology, the cosmic density decreases with time, and present massive objects 
are believed to have collapsed later than smaller ones. 

Recently, Prada et al. (2012, hearafter P12) determined $c(M_{\rm 200},z)$. 
According to their results, $c$ is described by a nearly universal U--shaped function of $M_{\rm 200}$, 
and the dependence on $z$ is more complex than proposed by Bullock et al. (2001). 
Ludlow et al. (2012) pointed out halos around the minimum of the U--shaped function 
are not relaxed dynamically and are in the early period of merger events. 
The analysis of P12 and Ludlow et al. (2012) was restricted to $M_{\rm 200} > 10^{11} M_{\rm \sun}$. 
In order to determine $c(M_{\rm 200},z)$ on less massive scales, we analysed the data of the cosmological 
simulation, {\it Cosmogrid}, performed by Ishiyama et al. (2013, hearafter I13). 
Though the covered volume is smaller than that of the Millenium simulations 
(Springel et al. 2005; Boylan-Kolchin et al. 2009), the Bolshoi (Klypin et al. 2011) and the MultiDark simulations (P12), 
the mass resolution is about 1000 times better (particle mass $\sim 10^5 M_{\rm \sun}$). 
Therefore we could analyse $c$ of DM halos down to $M \sim 10^8 M_{\rm \sun}$ which is comparable to the virial mass of dSphs. 

\begin{figure}
 \centering 
   \includegraphics[width=80mm, height=75mm]{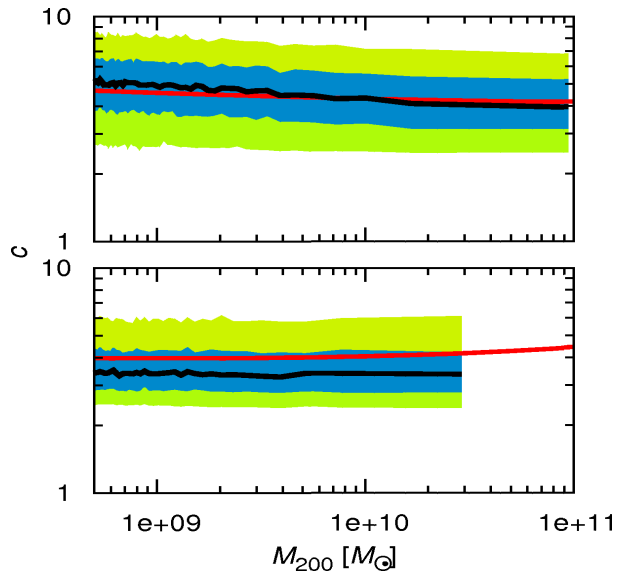}
     \caption{The concentration parameter, $c$, of DM halos as a function of the virial mass, $M_{\rm 200}$. 
         Red and black lines indicate the fitting formula $c(M_{\rm 200},z)$ proposed by Prada et al.
         (2012) and the median of our result, respectively. Blue and green areas
         cover halos distributed within the offsets of 25th and 45th percentiles
         from the median. Upper and lower panels correspond the results for $z=3.0$ and
         $z=5.4$, respectively. 
     }
  \label{c_parameter}
\end{figure}

Along the lines of P12, we computed $c$ of each halo from the velocity ratio $V_{\rm max}/V_{\rm 200}$. 
$V_{\rm max}$ and $V_{\rm 200}$ are the maximum circular velocity and the circular velocity at $r_{\rm 200}$,
respectively.
Fig. \ref{c_parameter} shows the concentration parameter of DM halos as a function of $M_{\rm 200}$.  
The fitting formula given by P12 perfectly predicts the result of our simulation at the redshift $z=3$, and reasonably matches that of the redshift $z=5.4$.
The small offset seen in the lower panel arise from the difference of the adopted cosmological parameters, especially $\sigma_{\rm 8}$. 
A similar level of offset is observed at $z=10.8$ (not shown). 
However, this is within a small error range and gives negligible effects on the following analysis.

\vspace{-5mm}
\section{The origin of the $\mu_{\rm 0D}$ relation}
Using the fitting formula $c(M, z)$, 
we study the connection between the cusp--to--core transformation and the $\mu_{\rm 0D}$ relation and estimate some characteristic properties of DM halos.
The procedures and assumptions of the analysis are as follows: 
\begin{itemize}
\item[(1)] Halo formation: We assume that DM halos form following an NFW profile with the $c(M_{\rm 200},z)$ of P12 for given $M_{\rm 200}$ and $z$. \\
\item[(2)] Cusp--to--Core transformation: We then assume that NFW halos are transformed into Burkert halos by 
the dynamical process quoted in \S1.
\end{itemize}

We then determine the parameters of the Burkert profile into which a given dark halo distribution will
transform. For that we impose two conditions. First, we require
mass conservation, that is, $M_{\rm 200}$ is conserved during the cusp--to--core transformation. 
Secondly, we assume that the density distribution is affected only in the inner regions while
the outer density distribution is preserved. This is a reasonable assumption if this transition is
a result of variations in the gravitational potential caused by baryons and stellar feedback.
From Equations (\ref{nfw}) and (\ref{burkert}), this condition requires 
\begin{equation}
\rho_{\rm s} r_{\rm s}^3 = \rho_0 r_0^3.
\label{outskirt}
\end{equation}

\begin{figure*}
 \centering 
   \includegraphics[width=120mm]{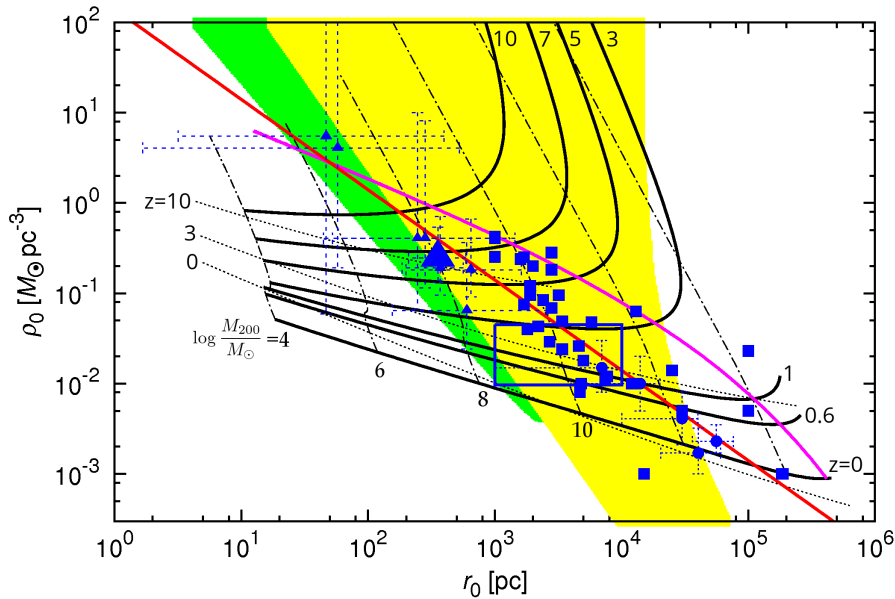}
     \caption{
              Diagram of $r_0$ vs. $\rho_0$. 
              Respective points show the parameters of Burkert profile, $\rho_0$ and $r_0$, obtained by observations. 
              Triangles, squares and circles represent the data from Salucci et al. (2012), Spano et al. (2008) and Donato et al. (2009), respectively. 
              Large triangle denotes the Sculptor dSph. 
              Red line is the scaling relation between $\rho_0$ and $r_0$ proposed by Salucci et al. (2012). 
              The place embraced by blue box corresponds to the results of Burkert (1995). 
              Solid and dotted black lines show the results of our analysis applying the $c(M,z)$ proposed by Prada et al. (2012) and Macci{\`o} et al. (2008), respectively. 
              Subscripts denote the corresponding transformation redshift, $z_{\rm t}$. 
              Black dot--dashed lines are the contour of halo mass. 
              Magenta line corresponds to the density peaks, $\delta = \delta_{\rm c} = 3 \sigma(M, z)$, where $\delta_{\rm c} = 1.69$ and $\sigma(M, z)$ are the critical density peak to collapse and the linear rms fluctuation of the density field for given mass and redshift, $\sigma(M,z)$, respectively. 
              Applying the Press--Schechter theory (Press \& Schechter 1974), most of halos satisfy $\delta < 3 \sigma(M, z)$ (below magenta line), stochastically. 
                The yellow and green regions indicate the parameter range satisfying the condition, $t_{\rm cool} \leq t_{\rm ff}$, where $t_{\rm cool}$ and $t_{\rm ff}$ are the cooling time and the free--fall time of the gas, respectively. 
                Aassuming a total gas mass equal to $0.1 \,M_{\rm 200}$, and a molecular abundance equal to 0.1\% of the neutral hydrogen, we estimate the cooling time using the cooling function of primordial gas given by Sutherland \& Dopita (1993, yellow) including ${\rm H_2}$ cooling (temperature below $\sim 10^4$ K, green) given by Galli \& Palla (1998).
     }
  \label{estimation}
\end{figure*}

We show the results in Fig. \ref{estimation} which plots $r_0$ vs. $\rho_0$.  
Though Spano et al. (2008) assumed an isothermal sphere to derive the core parameters, 
the difference between a Burkert profile and an isothermal sphere is small if the parameters 
(i.e., the central density and the core radius) are the same. 
Therefore we adopt the data from Spano et al. (2008) as parameters of the corresponding Burkert profile. 
The sample galaxies of Salucci et al. (2012, triangles) are nearby dSphs.
The data from Donato et al. (2009, circles) and Spano et al. (2008, squares) covers more massive galaxies. 
In Fig. \ref{estimation}, the yellow and green regions indicate that the free-fall time is longer than the cooling time assuming the collisional equilibrium of the primordial composition of the gas with hydrogen molecules. Such a parameter range, the cooling is efficient so that these objects can cool and condense leading to the active star formation.
Most of sample galaxies along the $\mu_{\rm 0D}$ relation lie in these regions. Therefore, we find that the scenario proposed in this study, the cusp--to--core transformation, naturally explains the $\mu_{\rm 0D}$ relation.

Figure \ref{estimation} also shows that the central density, $\rho_0$, is nearly constant for each cusp--to--core transformation redshift, $z_{\rm t}$, at which the central cusp has been transformed into core. 
We assume that DM halos have survived until now without significant change of the virial mass. 
Assuming that central cusps have been flattened out into cores by mechanisms which involve star formation activities (e.g., Ogiya \& Mori 2011, 2012), $z_{\rm t}$ corresponds to the formation epoch of DM halos and galaxies, approximately. 
As shown in Fig. \ref{c_parameter}, the dependence of $c$ on $M_{\rm 200}$ is weak, especially 
at high redshifts (see also Fig. 12 in P12).  Using this and the definition of the virial mass, Equation (\ref{virial_mass}), 
$r_{\rm s}^3 \propto M_{\rm 200}$ should be satisfied. 
We have defined the scale density, $\rho_{\rm s}$, by the mass profile of NFW profile, 
\begin{equation}
\rho_{\rm s} = M_{\rm 200} \bigl / \bigl [ 4 \pi r_{\rm s}^3 \bigl \{ \ln{(1+c)} - c/(1+c) \bigr \} \bigr ], 
\label{rhos}
\end{equation}
and found $\rho_{\rm s}$ is almost independent of $M_{\rm 200}$. 
As one of the results of the anasysis, we know that the core radius of Burkert halos, $r_0$, is close to the scale length of corresponding
NFW halos, $r_{\rm s}$. 
This means that $\rho_0 \approx \rho_{\rm s}$ since $\rho_0$ is determined by Equation (\ref{outskirt}). 
Combining everything, one can understand why the dependence of $\rho_0$ on $r_0$ is fairly weak. 
Fig. \ref{estimation} allows us to estimate $z_{\rm t}$ from $\rho_0$. 
As expected, the higher the core density, the higher the transformation redshift of the corresponding halo.
Halos with small core radii and correspondingly high core densities have small virial masses while large cores reside in massive dark halos.

Burkert (1995) obtained the density profiles of several DM halos. 
His sample halos had core radii of $10^3 {\rm pc} \la r_0 \la 10^4 {\rm pc}$. 
Fig. 1 in Burkert (1995) showed their outer density structures correspond to CDM halos formed at $z=0.6$. 
He also proposed a scaling relation between $r_0$ and $\rho_0$, and 
Salucci \& Burkert (2000) confirmed this relation for a larger sample. 
Our results match this relation (blue box). 
Salvadori et al. (2008) constructed a semi--analytical model to fit the observed features of the 
Sculptor dSph 
and found its transformation redshift to be about 7. 
Our core property 
estimate for this formation 
redshift matches nicely the observational measurements (large triangle). 
They also derived the virial mass of the Sculptor dSph which is about $10^8 M_{\rm \sun}$. 
Using the observationally inferred $r_0$ and transformation redshift, $z_{\rm t}$, 
a halo virial mass of $M_{\rm 200} \sim 1.2 \times 10^8 M_{\rm \sun}$ which is a perfect match to their result is derived.

As demonstrated in this section, the cusp--to--core transformation naturally generates 
the $\mu_{\rm 0D}$ relation in the context of the CDM cosmology. 
We also find that the cusp--to--core transformation redshift, $z_{\rm t}$, is closely related to the central density of cored DM halos.
Interestingly, the parameters of the Burkert profile ($r_0$ and $\rho_0$) are almost the same as the scaling parameters
of corresponding NFW halos ($r_{\rm s}$ and $\rho_{\rm s}$). 
Therefore we conclude that the observed universal correlation between the core parameters has been shaped
by the physics of formation of these DM halos, in other words, the cores of dark halos retain a memory of their formation
redshift.
The $c(M_{\rm 200}, z)$ relation is essentially important for these results since it determines the parameters of NFW profile. 

\vspace{-5mm}
\section{Relation between universalities}
The above mentioned $M(<300{\rm pc})$ universality has been reported for DM halos surround nearby dwarf galaxies. 
These galaxies are also included in the samples which were used to derive the $\mu_{\rm 0D}$ relation. 
In this section, we show the consistency between both relationships among nearby dwarfs 
and the inconsistency for larger and smaller galaxies. 
We start by adopting a Burkert profile for the mass--density structures of DM halos. 
The mass enclosed within given radius, $r$, is then
\begin{eqnarray}
M(r) = \pi \rho_0 r_0^3 \bigl [ -2 \arctan{( r/r_0)} +2 \ln{\bigl \{ 1+(r/r_0) \bigr \}} \nonumber \\
+ \ln{\bigl \{ 1+ (r/r_0)^2 \bigr \}} \bigr ]
\label{enclosed_mass}
\end{eqnarray}
(Mori \& Burkert 2000). 
We use Eqs. (\ref{rhos}) and (\ref{enclosed_mass}) in the analysis of this section.

\begin{figure}
 \centering 
   \includegraphics[width=80mm]{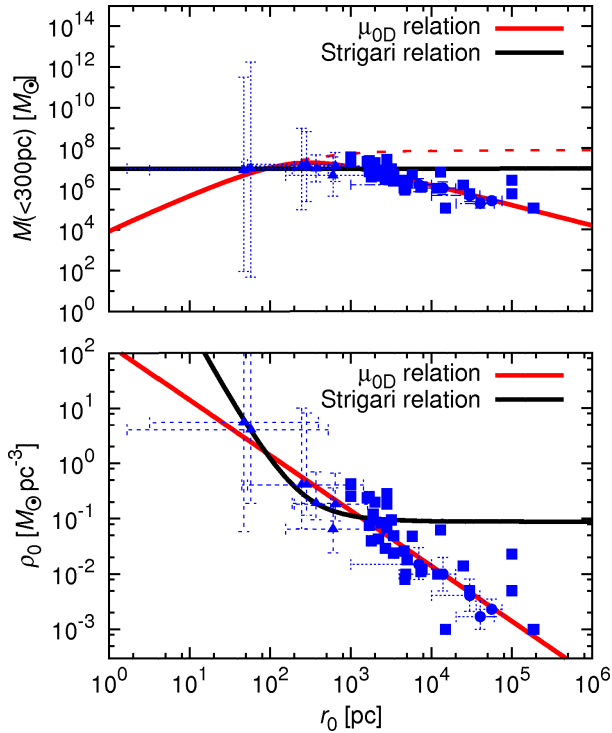}
     \caption{($Upper$) DM mass enclosed within 300pc, $M(<300{\rm pc})$, as a function of core radius, $r_0$. 
                        Symbols with error bars show the observations. 
                        The red, solid line depicts the $\mu_{\rm 0D}$ relation. 
                        We show $M(<300{\rm pc})$ of NFW halos with the assumption of 
                        $\rho_{\rm s} r_{\rm s} = 140 M_{\rm \sun} {\rm pc^{-2}}$ (red dashed).
                        The black line shows Strigari relation. 
              ($Lower$) The central density of Burkert profiles, $\rho_0$, 
                        is shown as a function of $r_0$. 
                        The black and red lines show Strigari and the $\mu_{\rm 0D}$ relation, respectively. 
                        Each symbol represents same observed galaxy within Fig. \ref{estimation}. 
}
  \label{universalities}
\end{figure}

The upper panel of Fig. \ref{universalities} shows the DM mass enclosed within 300pc, $M(<300{\rm pc})$, as a function of 
the core radius, $r_0$. 
As shown in this panel, the DM halos of dSphs have $\sim 10^7 M_{\rm \sun}$ within 300pc. 
This is consistent with Strigari relation (black line) for lower core radii. 
Independent of this study, Faerman et al. (2013) show Burkert profile naturally produces Strigari relation for nearby dwarf galaxies.  
However, halos of larger galaxies deviate from it. Our theoretical prediction, shown by the solid red line fits the data 
very well and indicates that Strigari relation is just a coincidence resulting from the fact that dwarf galaxies
like close to the maximum of $M(<300{\rm pc})$ versus $r_0$ correlation. 
As a result we would predict that
halos that are smaller than those typically found around dSphs should have smaller values of $M(<300{\rm pc})$ as 
predicted by Strigari relation. 
We also show $M(<300{\rm pc})$ of NFW halos supposing $\rho_{\rm s} r_{\rm s} = 140 M_{\rm \sun} {\rm pc^{-2}}$ (red dashed line). 
For this line, the horizontal axis corresponds to $r_{\rm s}$. 
Comparing the solid and dashed red lines one can see it is difficult to distinguish between
NFW halos and Burkert halos for dwarfs, but that measurements of core properties for more massive halos should
allow to distinguish between the different halo profiles.
The lower panel of Fig. \ref{universalities} shows the central density of the 
Burkert profile, $\rho_0$, as a function of $r_0$. 
This panel clearly demonstrates the two relations are consistent with each other among dSphs, 
but inconsistent for larger galaxies with scale radii of order 10 kpc or larger.

The consistency between Strigari and the $\mu_{\rm 0D}$ relation breaks down
for larger halos surround massive galaxies, but these galaxies follow the $\mu_{\rm 0D}$ relation very well. From that we
conclude that Strigari relation is just an indirect evidence of the $\mu_{\rm 0D}$ one for the case of
dwarf galaxies, since the latter covers a broader range of halo masses and halo core radii. 
In addition, the  NFW profile does not follow theses observations well, indicating that
the cusp--to--core transformation is an important process for the evolution of the inner dark halo structure.
 
\vspace{-5mm}
\section{Summary and discussion}
In this study, we have studied the relation between observational universalities of dark matter halos
and the cusp--to--core transformation. 
We obtained three interesting results: 
\begin{itemize}
\item[(1)] We verified the 
dependence of the concentration parameter of NFW profiles, 
$c$, on virial mass of halo, $M_{\rm 200}$, and redshift, $z$, proposed by Prada et al. (2012) using the data of the 
cosmological simulation performed by Ishiyama et al. (2013). 
According to our analysis, the $c(M_{\rm 200},z)$ proposed by Prada et al. (2012) is appropriate down to 
$M \sim 10^8 M_{\rm \sun}$ which is the mass scale of dwarf galaxies. \\
\item[(2)] The cusp--to--core transformation shapes the $\mu_{\rm 0D}$ relation naturally. 
The central density of the Burkert profile, $\rho_0$, is almost independent of $M_{\rm 200}$. 
We therefore can estimate the formation epoch of the respective halos from $\rho_0$. 
Our estimations are consistent with previous studies. 
The $\mu_{\rm 0D}$ relation 
is a relict of the formation phase of DM halos and keeps a memory of their time of formation. \\
\item[(3)] We found that the $\mu_{\rm 0D}$ and Strigari relations are consistent with each other among DM halos surrounding dSphs. 
However, this consistency breaks down for halos of more massive galaxies that have larger cores. 
The $\mu_{\rm 0D}$ relation can be applied to a broader range of halo mass than Strigari relation. 
We conclude that Strigari relation is an indirect evidence of the $\mu_{\rm 0D}$ relation. 
\end{itemize}

The cusp--to--core transformation represents a viable model in order to explain the origin of these universalities.
In our analyses, we assumed a transformation from NFW profiles to Burkert profiles. 
This assumption requires some mechanisms to flatten the cusps and to create the cores
with sizes comparable to the scale lengths of the cuspy halo. 
This requirement might allow to distinguish between different models of core formation and might provide important constraints for them.
It is not only the flattening of a cusp but also the formation of a reasonably sized core that
these models have to generate. 

An interesting question is whether the cusp--to--core transformation also solves the missing satellite (e.g., Moore et al. 1999b) problem.
This might be difficult to achieve as obviously the cores are observed in dwarf galaxies that were not disrupted by internal
processes or that became invisible because of quenching of star formation but that still experienced major
changes in the gravitational potential leading to cores (see also Pe{\~n}arrubia et al. 2012).
In addition, it is also intersting about a weak dependence of a halo mass on the $\mu_{\rm 0D}$ relation (Cardone \& Tortora 2010; Boyarsky et al. 2010; Cardone \& Del Popolo 2012). 
The subsequent studies will report on these issues.

It is noteworthy that the transformation redshifts of the Sextans and the Leo II dSphs, the leftmost points in Fig. 2, are beyond the redshift $z_{\rm t}= 10$. 
In such an early epoch, the gas is almost metal--free, and ${\rm H_2}$ acts as a prime coolant (e.g., Galli \& Palla 1998; Omukai \& Nishi 1998; Bromm \& Yoshida 2011 and references therein). 
Further studies about these galaxies may play a key role to unravel the formation of the first objects in the universe.

\vspace{-5mm}
\section*{Acknowledgments}
We are grateful to the anonymous referee for providing many helpful comments and suggestions. 
This work was supported in part by JSPS Grants-in-Aid for Scientific Research: (A) (21244013), (C) (25400222), (S) (20224002), and (B) (24740115), Grants-in-Aid for Specially Promoted Research by MEXT (16002003), and Grant-in-Aid for JSPS Fellows (25--1455 GO). AB acknowledges support from the cluster of excellence "Origin and Structure of the Universe". TI is financially supported by MEXT HPCI STRATEGIC PROGRAM.

\label{lastpage}
\end{document}